# Nonreciprocal field transformation with active acoustic metasurfaces

Xinhua Wen[1], Choonlae Cho[2], Xinghong Zhu[1], Namkyoo Park[2]*, Jensen Li[1]*

[1]*Department of Physics, The Hong Kong University of Science and Technology, Kowloon, Hong Kong, China.*

[2]*Photonic Systems Laboratory, Department of Electrical and Computer Engineering, Seoul National University, Seoul 08826, South Korea*

Field transformation, as an extension of the transformation optics, provides a unique means for nonreciprocal wave manipulation, while the experimental realization remains a significant challenge as it requires stringent material parameters of the metamaterials, e.g., purely nonreciprocal bianisotropic parameters. Here, we develop and demonstrate a nonreciprocal field transformation in a 2D acoustic system, using an active metasurface that can independently control all constitutive parameters and achieve purely nonreciprocal Willis coupling. The field-transforming metasurface enables tailor-made field distribution manipulation, achieving localized field amplification by a predetermined ratio. Interestingly, the metasurface demonstrates the self-adaptive capability to various excitation conditions and can extend to other geometric shapes. The metasurface also achieves nonreciprocal wave propagation for internal and external excitations, demonstrating a one-way acoustic device. Such a field-transforming metasurface not only extends the framework of the transformation theory for nonreciprocal wave manipulation, but also holds significant potential in applications such as ultra-sensitive sensors and nonreciprocal communication.

Over the past two decades, metamaterials with engineered material parameters have provided a remarkable way to achieve wave manipulation for different classical waves like electromagnetic (*1-4*), acoustic (*5-8*) and elastic waves (*9-12*). The combination of metamaterials and the concept of transformation optics allows us to engineer the wave propagation driven by coordinate transformations (*13-16*), enabling the concepts of cloaking and illusion that appeared in science fiction to become a reality. The concept of transformation optics is later extended to acoustic and elastic waves, and many novel devices designed by the transformation optics approach, including invisibility cloaks (*17-20*), illusion devices (*21,22*), wave concentrators (*23,24*), and rotators (*25,26*) have been demonstrated. The vast majority of these devices intrinsically conform to the principle of reciprocity, as they are designed based on the transformation that uses the metric tensor independent of direction (*13,14*). Nonetheless, controlling nonreciprocal wave propagation using the transformation approach, like nonreciprocal cloaking and sensing (*27,28*), remains highly desirable for extending the transformation approach. It has been proposed that nonreciprocal cloaking can be achieved with a coordinate-transformed gyrotropic medium in an external magnetic field (*27*). However, to attain magnetic-free nonreciprocity, the framework of conventional transformation optics rooted in Lorentz reciprocity is not easily applicable. An extension of transformation optics that combines frame deformation and coordinate transformation has been recently proposed for an electromagnetic continuum with local rotation at each point, to obtain a nonreciprocal response (*29*), while the implementation of such a structural anisotropy with metamaterials is very challenging.

On the other hand, field transformation (*30-33*), as another extension of the transformation approach, transforms the fields directly instead of using coordinate transformation, offering a unique means for nonreciprocal wave manipulation. By choosing a scaling-type transformation function, i.e., directly scaling the complex amplitudes of the fields, one can obtain nonreciprocal constitutive relations. Such a particular kind of field transformation can be exploited to realize nonreciprocal field transformation, with a field-transforming metamaterial that has the form of material relations of slowly moving media (*30,34*). To satisfy such a material relation, it requests a metamaterial that can flexibly and independently control all constitutive parameters, and also generate purely nonreciprocal bianisotropic coupling parameters. This is quite challenging for passive metamaterials with the constraint of reciprocity, even for conventional active metamaterials (*35 - 37*), thus the experimental realization of nonreciprocal field transformation has never been demonstrated.

In acoustics, active meta-atoms with a software-defined impulse response (*38-40*), allow us to independently control all constitutive parameters and achieve nonreciprocal acoustic bianisotropic coupling, also known as Willis coupling (*41 - 44*), opening the possibility for the realization of nonreciprocal field transformation. In this work, we propose and experimentally demonstrate nonreciprocal field transformation in a 2D acoustic system. Our field-transforming metasurface constructed by active meta-atoms, allows to locally amplify the field within a specific region without phase distortion, while the field outside is not affected due to the elimination of scattering waves. In contrast to conventional active control systems that require a priori knowledge of the incident wave field or sufficient latency time for recalculating the required secondary sources (*45,46*), here the field-transforming metasurface can already adapt to various incident waves with a tailor-made and fixed material response. Furthermore, the metasurface also enables a one-way acoustic device with nonreciprocal transmission in response to internal and external excitations. Nonreciprocal field transformation significantly extends the framework of transformation theory for nonreciprocal wave manipulation, but also holds great potential for achieving highly sensitive sensing and nonreciprocal communication.

## Nonreciprocal metamaterial approach for acoustic field transformation

Field-transforming metamaterials allow for desired field distribution within the medium. Here we demonstrate the acoustic wave manipulation using a field-transforming metasurface, for convenience, with cylindrical symmetry. We start from the desired field distribution using a ring-shaped field-transforming metasurface with inner and outer radii $R_1$ and $R_2$, as illustrated in Fig. 1A. For an external incident wave outside the metasurface, the metasurface transforms the pressure field by a transformation function $f(\boldsymbol{r})$, resulting that the pressure field within the region enclosed by the metasurface (dark blue region) is amplified by a factor of $1 + s$, i.e., $p(\boldsymbol{r}) = (1 + s)p_0(\boldsymbol{r})$, where $p_0(\boldsymbol{r})$ represents the propagating pressure field in the background, and a positive $s$ denotes field amplification. On the other hand, we aim for the pressure field outside the metasurface (cyan region) to remain unaffected, ensuring $p(r) = p_0(r)$ by eliminating the scattered field outside the metasurface (to prevent backscattering to the source). Such a field manipulation is illustrated by the radial mapping of the pressure field along the red dashed line, as shown in the lower panel of Fig. 1A. The field exhibits a scaling factor of $1 + s$, gradually decreasing (from $R_1$ to $R_2$) back to a value of 1 at the outer boundary of the metasurface. The transformation function $f(r)$ can be flexibly designed depending on the practical applications.

To achieve the desired field manipulation, we employ an active metasurface with the necessary material parameters to be deduced from a theory of field transformation. As shown in Fig. 1B, the ring-shaped metasurface comprises 24 active meta-atoms, with a distance of $\Delta l = 0.039$ m between neighboring atoms (i.e., the center circle of the metasurface has a radius of 0.15 m). Each meta-atom, denoted by a yellow dashed rectangle, is shown in the photographs in Fig. 1C. It consists of two microphones ($D_1$ and $D_2$) and two speakers ($S_1$ and $S_2$), integrated into a PCB board with a size of $0.03 \times 0.06$ m. Such PCB board interconnects to a microcontroller (Itsbitsy M4), referred to as a digitally virtualized atom (*38-40*), allowing to implement a customized scattering response as four convolution kernels $Y_{ij}$ in the time domain, as schematically shown with the yellow arrows in Fig. 1D. For the experiment, meta-atoms are affixed to the bottom plate (cyan color) of a 2D waveguide, with the microphones and speakers facing upwards to sense and re-radiate into the waveguide. Four measurement microphones (labeled as $M_1$ to $M_4$), inserted into the top plate (purple color) that connects to a movable positioning stage, enable the scanning of the pressure field across the entire region.

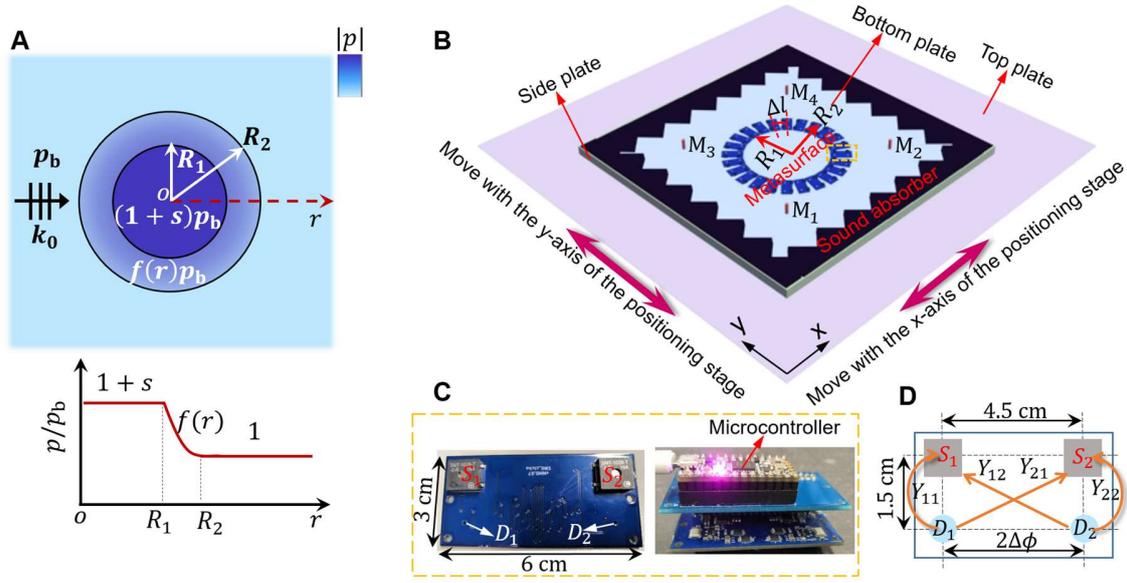

**Fig. 1. Localized field amplification and its implementation setup.** (**A**) Schematic of the localized field amplification by a ring-shaped metasurface with inner and outer radii $R_1$ and $R_2$ (upper panel). The pressure field inside the metasurface (dark blue region) is amplified by a factor of $1 + s$, while the pressure field outside (cyan region) remains unaffected. Lower panel: the radial mapping of the pressure field along the red dashed line. (**B**) Experiment setup for 2D field pattern scanning. A metasurface consisting of 24 active meta-atoms is affixed to the bottom plate (cyan color) of a 2D waveguide. The top plate of the waveguide (purple color), with four inserted microphones (labeled as $M_1$ to $M_4$), is connected to a movable positioning stage to enable scanning across the entire region. (**C**) Photographs of the meta-atom. Two microphones ($D_1$ and $D_2$) and two speakers ($S_1$ and $S_2$) are integrated into a PCB board, and interconnected to a microcontroller. (**D**) Schematic representation of the meta-atom that performs four convolution kernels $Y_{ij}$ in the time domain, indicated by orange arrows.

To deduce the required material parameters for implementation, we employ a theory of field transformation without any coordinate transformation (*30*). The acoustic wave equations in cylindrical coordinates $(r, \theta)$, at a harmonic radial frequency $\omega$ (with $e^{-i\omega t}$ convention), can be expressed as

$$\frac{\partial p}{\partial \theta} = i\omega r(\rho_0 v_\theta + d_\theta), \qquad \frac{\partial p}{\partial r} = i\omega(\rho_0 v_r + d_r)$$

$$\frac{\partial r v_r}{\partial r} + \frac{\partial v_\theta}{\partial \theta} = i\omega r(\beta_0 p + m). \tag{1}$$

Here, $p$ represents the pressure field, $v_r$ and $v_\theta$ are the radial and azimuthal components of the velocity field, respectively. $m$ corresponds to the acoustic monopole density generated by the metasurface, while $d_r$ and $d_\theta$ represent the radial and azimuthal components of the acoustic dipole moment density. $\beta_0$ and $\rho_0$ denote the free-space compressibility and density, respectively. To manipulate the local field amplitudes, we request a scaling-type field transformation given by

$$p = f(r)p_0, \qquad v_r = f(r)v_{r0}, \qquad v_\theta = f(r)v_{\theta 0}, \tag{2}$$

where $f(r)$ is a transformation function (only in terms) of $r$ to relate the original (without the presence of the field-transforming metasurface) and the transformed fields. The background pressure field $p_0$, and velocity field $v_{r0}$ and $v_{\theta 0}$ represent the background fields that satisfy the free-space wave equation, i.e., Eq. (1) with zero acoustic monopole and dipole moments. By substituting Eq. (2) into Eq. (1), we obtain the required monopole and dipole moments

$$m = \frac{f'(r)}{i\omega}v_{r0}, \qquad d_r = \frac{f'(r)}{i\omega}p_0, \qquad d_\theta = 0. \tag{3}$$

This is akin to those active control systems where the induced secondary sources are determined by the incident wave (45). Thus, the secondary sources need to be recalculated when the incident field changes. However, by combining Eq. (2), Eq. (3) can be rewritten as

$$m = \frac{f'(r)}{i\omega f(r)}v_r, \qquad d_r = \frac{f'(r)}{i\omega f(r)}p, \qquad d_\theta = 0. \tag{4}$$

For our field-transforming metasurface, the secondary source can be associated with the local field directly. As a result, the active metasurface with a fixed effective material response, can adapt to the various incident fields without prior knowledge of the incident wavefield. Moreover, the fixed material response can be specified by controlling the convolution kernels in the program, which will be discussed in detail later. Furthermore, Eq. (4) can be expressed as a dimensionless constitutive matrix relating the volume strain $-\epsilon$ and momentum $\boldsymbol{\mu}$ to the pressure and velocity fields (42)

$$\begin{pmatrix} -\epsilon/\beta_0 \\ c\mu_r \\ c\mu_\theta \end{pmatrix} = \begin{pmatrix} \beta & i\tau & 0 \\ i\tau' & \rho & 0 \\ 0 & 0 & \rho \end{pmatrix} \begin{pmatrix} p \\ \eta_0 v_r \\ \eta_0 v_\theta \end{pmatrix} \qquad (5)$$

with compressibility and density $\beta = \rho = 1$ and the Willis couplings $\tau = \tau' = -f'(r)/(k_0 f(r))$. $k_0 = \omega/c$ is the free-space wavenumber, and $c$ represents the sound speed in free space. $\eta_0 = \rho_0 c$ corresponds to the wave impedance in free space. The constitutive matrix of the field-transforming metasurface in Eq. (5), within the region with $\tau = \tau' \neq 0$, is symmetric (purely nonreciprocal) rather than anti-symmetric ($\tau = -\tau'$, reciprocal) and is simultaneously non-Hermitian (active) (40, 47). Additionally, $\beta = \rho = 1$, indicates that the metasurface possesses the same compressibility and density as that of free space (i.e., $\beta = \beta_0, \rho = \rho_0$). In fact, the wave impedance of the metasurface does not change despite the presence of the purely nonreciprocal Willis coupling $\tau$, since the pressure and velocity fields are transformed in the same way (30). Consequently, the scattered field outside the metasurface is eliminated, leaving the pressure field outside intact.

By choosing an appropriate transformation function $f(r)$, we can obtain the required constitutive parameters of the ring-shaped metasurface. At the inner and outer boundaries, the pressure fields experience different scaling factors defined as $f(R_1) = f_1$ and $f(R_2) = f_2$. Taking $\ln f$ to be linear in $r$, then the transformation function (for $R_1 \leq r \leq R_2$) can be written as

$$\ln f = \frac{R_2 - r}{R_2 - R_1} \ln f_1 + \frac{r - R_1}{R_2 - R_1} \ln f_2, \qquad (6)$$

Thus the nonreciprocal Willis coupling $\tau$ can be obtained as

$$\tau = -\frac{f'(r)}{k_0 f(r)} = -\frac{\ln f_2 - \ln f_1}{k_0 (R_2 - R_1)}. \qquad (7)$$

Unlike the conventional transformation optics approach that typically requires a metamaterial with an inhomogeneous material profile to achieve the desired field manipulation, the field-transforming metasurface described here is purposely designed to have homogenous material parameters. This is achieved by setting the linear property of

$\ln f$ with respect to $r$, making the experimental implementation much simpler. According to Eq. (7), the local field amplification ratio can be equivalently written as

$$f_1/f_2 = e^{\tau k_0 (R_2 - R_1)} = 1 + s, \qquad (8)$$

Therefore, a positive real nonreciprocal Willis coupling $\tau$ enables filed amplitude amplification without phase distortion, which holds significant potential in the application for highly sensitive sensors. Moreover, the amplification ratio $1 + s$ can be adjusted by controlling the value of $\tau$.

Before delving into the implementation of the metasurface, we first verify our theoretical framework of field transformation using full-wave simulations at the level of material parameters derived above. As an example, we set the local field amplification ratio to be $1 + s = 1.5$ (i.e., $s = 0.5$) throughout the main text without further mention. We set the constitutive parameters of the metasurface: $\tau = 0.27$ with $\beta = \rho = 1$ and a thickness of metasurface as 0.24 of the free-space wavelength $\lambda$ (i.e. $R_2 - R_1 = 0.24\lambda$) in order to satisfy Eq. (8). By specifying these material parameters for the field-transforming metasurface in COMSOL Multiphysics (general PDE module), we can simulate the local field amplification phenomenon. As shown in Fig. 2B, the field-transforming metasurface enables the amplification of the pressure field inside the metasurface without any phase distortion, compared to the situation without the metasurface present (Fig. 2A). Furthermore, the pressure field outside the metasurface is unaffected. Figure 2C shows the radial mapping of the nonreciprocal Willis coupling $\tau$ (upper) and the pressure field amplitude $|p|$ (lower) along the white lines in Fig. 2(A and B), verifying that the ring-shaped metasurface with a homogenous $\tau = 0.27$ (red line) indeed enables the local field amplification inside the metasurface ($x/\lambda \leq R_1$) with a ratio of 1.5 (purple line). In addition, the pressure field outside the metasurface ($x/\lambda \geq R_2$) equals the background pressure field (purple and green dashed lines), demonstrating the complete suppression of the scattered field outside due to impedance matching. Consequently, such a metasurface indeed provides the desired field distribution in Fig. 1A.

Since the generated secondary source of the metasurface can be directly associated with the local field (Eq. (4)), the field-transforming metasurface can adapt to other types of external incident waves, like the circular wavefront from a point source. As shown in Fig. 2(D and E), the desired local field amplification also occurs for an external point source excitation at $(5\lambda, -4\lambda)$. Outside the metasurface, the total field remains the same as the original cylindrical wave (i.e., the metasurface is absent) without scattering. In addition, the shape of the field-transforming metasurface is not limited to the geometry shape with rotational symmetry, and can also be extended to other shapes, like the bean shape shown in Fig. 2F. The bean-shaped metasurface with an equal distance between the inner and outer boundaries ($0.24\lambda$), has the same constitutive parameters as that of the circular one, also allows for the desired local field amplification.

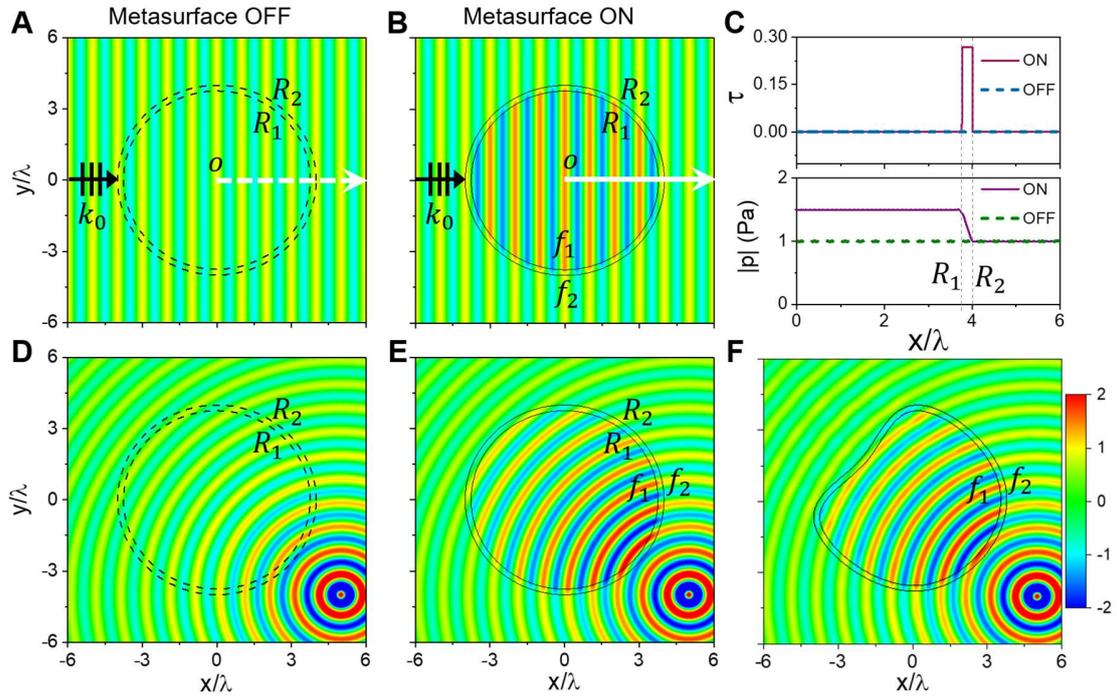

**Fig. 2. Numerical verification of the field-transforming metasurface with the required material parameters**. (**A** and **B**) Simulated pressure field patterns for an external plane wave excitation when the metasurface is (**A**) absent (denoted as OFF) and (**B**) present (denoted as ON). (**C**) The radial mapping of the Willis coupling $\tau$ (upper) and pressure field amplitude (lower) along the white lines in (**A** and **B**). (**D** and **E**) Similar to (**A** and **B**) but for an external point source excitation at $(5\lambda, -4\lambda)$. (**F**) The simulated pressure field pattern for a bean-shaped metasurface with the same constitutive parameters. In (**A** to **E**), $R_1 = 3.76 \lambda$, $R_2 = 4.0 \lambda$ with $\lambda$ being the free-space wavelength. $f_1$ ($f_2$) corresponds to the scaling factor at the inner (outer) boundary of the metasurface.

## Experimental verification

For the experimental realization, we turn to a metasurface constructed with 24 active meta-atoms (Fig. 1B). By manipulating the software-defined convolution kernels $Y_{ij}$ of the meta-atoms, all constitutive parameters of the metasurface, including compressibility, density and the Willis coupling can be independently controlled (*38-40*). Particularly, such an active metasurface enables the realization of purely nonreciprocal Willis coupling, in contrast to those typical Willis metamaterials that can only generate reciprocal Willis coupling (*42-44*). Therefore, a field-transforming metasurface with required constitutive parameters (Eqs. (5) and (7)) can be experimentally realized, by controlling four time-convolution kernels $Y_{ij}$ of the meta-atoms that connect the detected pressure fields by two microphones to the radiated pressure fields from two speakers (denoted by yellow arrows in Fig. 1D). In time harmonic, the time convolution in each meta-atom can be expressed as a matrix multiplication as:

$$\begin{pmatrix} S_1 \\ S_2 \end{pmatrix} = \begin{pmatrix} Y_{11} & Y_{12} \\ Y_{21} & Y_{22} \end{pmatrix} \begin{pmatrix} D_1 \\ D_2 \end{pmatrix} \tag{9}$$

For our field-transforming metasurface with $\beta = \rho = 1$, and nonreciprocal Willis coupling $\tau$ from Eq. (7), the convolution kernels $Y_{ij}$ can be set as (see derivation in Note S1)

$$Y_{11} = Y_{22} = 0,$$

$$Y_{12} = \frac{1}{2}i(-1 + e^{\tau\phi_0})\csc(2\Delta\phi), \qquad Y_{21} = \frac{1}{2}i(-1 + e^{-\tau\phi_0})\csc(2\Delta\phi). \tag{10}$$

where $\phi_0 = k_0(R_2 - R_1)$ is the phase that elapses across the air with the same thickness of the metasurface in the radial direction, and $\Delta\phi$ is the propagation phase distance from the center of the meta-atom to either one of the speakers or microphones (Fig. 1D). Such convolution kernels in Eq. (10) allow the suppression of the backscattering fields and the realization of nonreciprocal Willis coupling due to $Y_{12} \neq Y_{21}$.

In the experiment, the constructed ring-shaped metasurface has a thickness of $R_2 - R_1 = 0.06$ m (see Fig. 1(B and C)), and the incident wavelength is chosen as $\lambda = 0.25$ m (i.e., working frequency is 1372 Hz), to obtain the amplification ratio 1.5 (satisfying Eq.

(8)). To achieve the required nonreciprocal Willis coupling $\tau = 0.27$, the necessary convolution kernels $Y_{12}$ and $Y_{21}$ of all meta-atoms can be obtained from Eq. (10)

$$Y_{12} = 0.276i, Y_{21} = -0.184i \qquad (11)$$

As resonance is helpful to greatly enhance Willis coupling (*47*) and ensure our system stability (*40*), we employ Lorentzian-type resonance for the convolution kernels $Y_{12}$ and $Y_{21}$ in the implementation:

$$Y_{12/21}(f) = \frac{g_{12/21} f_0}{f_0^2 - (f + i\gamma)^2}. \qquad (12)$$

Here, the resonance linewidth $\gamma$ of 50 Hz, and the resonance frequency $f_0$ is chosen as the working frequency of the metasurface 1372 Hz, such that $Y_{12/21}$ can be approximated as $Y_{12/21}(f_0) \cong ig_{12/21}/2\gamma$, a purely imaginary number at the resonance frequency. Then we can control the resonance strength $g_{12/21}$ to obtain the specified convolution kernels in Eq. (11) at $f_0$. In the actual implementation, the radiated pressure field from two speakers is associated with the distance between neighboring meta-atoms $\Delta l$, thus we first implement and test the meta-atom in a 1D waveguide, to verify that the metasurface has the required constitutive parameters (see details in Note S2).

Using the field-transforming metasurface with the required constitutive parameters, we now experimentally validate the localized field amplification. In the experiment, the metasurface features a central circle with a radius of 0.15 m, which is henceforth represented as a black circular line. As depicted in Fig. 3B, when a point source excitation at 1372 Hz is applied at the point (0.29 m, -0.1 m), labeled as point A, the measured pressure field within the region enclosed by the metasurface is significantly amplified compared to the case in Fig. 3A when the metasurface is turned off. The amplification ratio will be discussed later. The metasurface, featuring purely nonreciprocal Willis coupling and impedance matching, only affects the field pattern enclosed by the metasurface. Essentially, the secondary sources of all the meta-atoms in the metasurface interfere to generate a scattered field distributed only inside the metasurface, propagating in-phase with the background field (refer to Note S3 for further details). Thus, constructive interference between the scattered field and background field occurs, leading to local field amplification inside the metasurface. Meanwhile, the scattered field outside is almost completely suppressed, leaving the background field outside unaffected.

With a fixed material response defined in the program, the active metasurface can adapt to different incident waves automatically without the need for reconfiguration of the meta-atoms. As illustrated in Fig. 3(C and D), the localized field amplification also occurs when the metasurface is activated (Fig. 3D) for a point source excitation at a different location A' (-0.27 m, 0.18 m). In this case, the scattered field outside is also suppressed (see Note S3), although some nonvanishing scattered fields, resulting from the imperfect sound absorber, are present around the boundaries of the experimental domain (Fig. 3(C and D)). The localized field amplification and the self-adaptive property can also be numerically verified in COMSOL Multiphysics using a metasurface with 24 discrete meta-atoms, as discussed in Note S4. Numerical results also confirm that the field-transforming metasurface can adapt to external point source excitations at different locations, and a plane wave excitation with an arbitrary incidence angle (Note S4).

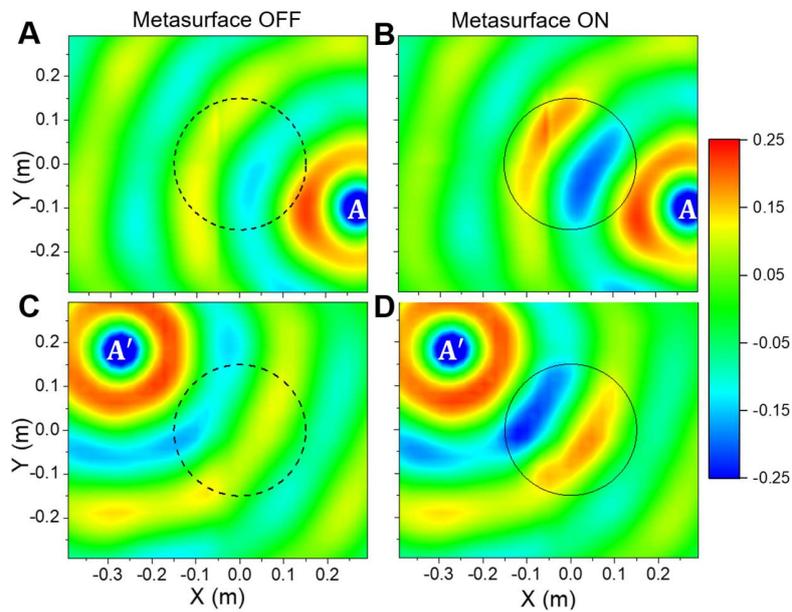

**Fig. 3. Experimental demonstration of localized field amplification.** (**A** and **B**) Measured pressure field pattern for a point source at location A when the metasurface is (**A**) turned off and (**B**) turned on. The solid (dashed) black circular line represents the turned-on (turned-off) metasurface with a ring shape. (**C** and **D**) Similar to in (**A** and **B**) but for a point source excitation at location A'.

The same field-transforming metasurface can also be used to achieve a one-way acoustic device with nonreciprocal transmission. As schematically shown in Fig. 4A, for an external point source excitation at point A, the transmitted wave detected at point B (0.09 m, 0 m) is amplified by a factor of $1 + s$ (indicated by the red arrow). Conversely, for a point source excitation at point B, within the region enclosed by the metasurface, the transmitted wave detected at point A is scaled by a factor of $1/(1 + s)$ (blue arrow), showcasing nonreciprocal wave propagation. Figure 4B displays the simulated pressure field pattern for a point source excitation at point B, utilizing the metasurface with 24 discrete active meta-atoms (see Note S4 for the simulation approach). The pressure field outside the metasurface is significantly reduced due to the destructive interference between the scattered field and the background field (see Note S3). However, the pressure field within the region enclosed by the metasurfaces remains unaffected, due to the suppression of backscattering signals. The measured field pattern in Fig. 4C closely matches the simulation result (Fig. 4B), providing experimental validation of nonreciprocal transmission using our field-transforming metasurface. Therefore, we experimentally demonstrate a one-way device that allows a sensor inside to detect the sound from outside clearly, but not vice versa.

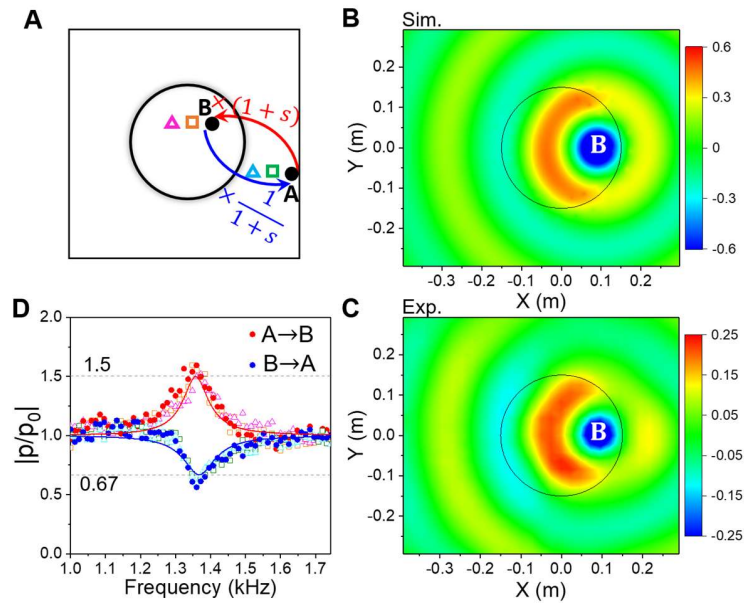

**Fig. 4. Nonreciprocal wave propagation with the field-transforming metasurface.** (**A**) Schematic of the nonreciprocal transmission for external and internal point source excitations. The transmitted wave is scaled

by a factor of $1 + s$ from A to B (external excitation, red arrow), while the transmitted wave is scaled by a factor of $1/(1 + s)$ from B to A (internal excitation, blue arrow). (**B**) Simulated and (**C**) measured pressure field pattern for an internal point source at location B. (**D**) The field scaling factor $|p/p_0|$ as a function of frequency. Symbols represent the measured results, and lines represent the model results.

For the nonreciprocal transmission, we experimentally validate the field scaling factors for two opposite propagation directions. As shown in Fig. 4A, we place a point source at point A (B) and measure the pressure at point B (A) and two nearby points denoted by open square and triangular symbols (all three measured points have an equal distance of 2.54 cm). The scaling factor is defined as the pressure field amplitude ratio with the metasurface turned on and off ($|p/p_0|$). Figure 4D presents the measured scaling factor in symbols as a function of incident frequency, where the lines represent the model results calculated from Eq. (8) and Eq. (12) (with frequency shift). At the working frequency of 1372 Hz, the detected pressure field at point B (A) is indeed scaled by a ratio of around 1.5 (0.61) for a point source excitation at point A (B), which aligns with the expected values indicated by the dashed gray lines. The pressure field detected at the two nearby points (open square and triangular symbols) also shows consistent experimental results. Although the spectrum demonstrates a limited working bandwidth (with a FWHM of 87 Hz) due to the employed resonance (Eq. (12)), this is helpful for the stability of a system with gain, as it effectively precludes the amplification of white noise beyond the desired frequency range. Moreover, the active metasurface, endowed with programmability, allows for the reconfigurability of the working bandwidth and frequency (*38*). As previously mentioned, the field-transforming metasurface can extend to other geometry shapes (Fig. 2F), and we also numerically verify nonreciprocal wave propagation using a bean-shaped metasurface comprising 24 discrete meta-atoms (see Note S5). Additionally, a larger amplification ratio can be achieved by increasing the nonreciprocal Willis coupling, as discussed in Note S6.

## DISCUSSION

We proposed and experimentally demonstrated nonreciprocal field transformation in a 2D acoustic system. Using active meta-atoms with a software-defined material response, we realize a metasurface with the purely nonreciprocal Willis coupling terms and the same compressibility and density as that of free space (eq. (5)). The field-transforming metasurface enables tailor-made field distribution manipulation, achieving the localized field amplification by a predesigned ratio and nonreciprocal wave propagation. Our scale-type intensity manipulation approach stands in stark contrast to those field intensity manipulation approaches that rely on a judiciously designed gain-loss profile (*48-51*). The latter remains constrained by reciprocity and achieves the desired field manipulation only for a specific incident direction, while our approach breaks reciprocity and demonstrates a self-adaptive capability to various incident conditions, either a plane wave excitation with an arbitrary incident angle or a point source excitation. In addition, we present the first experimental demonstration of nonreciprocal wave control using the transformation approach, significantly extending the framework of the transformation theory for nonreciprocal wave manipulation and anticipating to trigger more nonreciprocal devices designed by transformation approach like nonreciprocal cloaking device (*28*). Furthermore, the field-transforming metasurface provides great potential for various applications, like ultra-sensitive sensors and nonreciprocal communication. For example, the field-transforming metasurface substantially enhances sensor sensitivity while suppressing backscattering waves without disturbing the external field. Such a highly sensitive sensor can adapt to various incident waves as needed, even in a dynamic environment.

## METHODS

The meta-atom consists of two microphones (ADMP401) and two speakers (SMT-1028-T-2-R) that are integrated into a PCB board (with a size of 0.03×0.06 m), connected to a microcontroller (Adafruit ItsyBitsy M4 Express). The microcontroller performs four different channels of time convolution in connecting the two speakers. In the actual experiment, the efficiency of the speakers that radiate the pressure field converted from the volumetric flow depends on the type of speakers used in the experiment. Therefore, calibration factors in the program are used to obtain the expected convolution kernels when implementing the meta-atom in the 1D waveguide. Once the implemented program for the meta-atom in the 1D waveguide is tested to achieve the required constitutive parameters, we upload the same program for all the meta-atoms of the metasurface.

As shown in Fig. 1B, the whole 2D experimental setup consists of a ring-shaped metasurface, a 2D waveguide made of acrylic material, a movable positioning stage (not shown) and the pressure field measurement system with four microphones connecting to the NI DAQ. The meta-atoms of the metasurface are affixed in the bottom plate (cyan color) of the box, thus not blocking the wave propagation inside the waveguide. Specifically, there are preserved holes on the bottom plate for microphones and speakers, to sense and re-radiate the pressure field into the 2D waveguide. Sound absorbers made of acoustic foams are placed around the four rigid side plates. Four microphones labeled as $M_1$-$M_4$, are inserted in the preserved holes on the top plate (purple color) to measure the pressure field inside the waveguide. For pressure field pattern scanning, the top cover connects to a movable positioning stage, such that a maximum area of 0.693 × 0.586 m can be scanned by moving the top plate. And we use a Labview system to control the positioning stage and take the time data from the NI DAQ device connecting to four microphones. Using the Fourier transform, we can get the frequency data in every scanned position, obtaining the pressure field pattern. It is worth noting that the height of the 2D waveguide is 0.02 m, thus only the fundamental waveguide mode for the working frequency 1372 Hz is allowed to propagate inside the box, which guarantees an experimental system of a 2D nature.

**Acknowledgments**

The authors thank Prof. Shuang Zhang for useful discussions. X.W. would like to thank Jusung Park, Hansol Noh and Gyoungsub Yoon for their help when she was in SNU for this project. J.L. acknowledges support from Research Grants Council (RGC) of Hong Kong through projects no. 16303019, 16307522, and AoE/P-502/20 and the Croucher Foundation (CF23SC01). N.P. acknowledges support from National Research Foundation of Korea (NRF) through the Global Frontier Program (No. 2014M3A6B3063708).

**Author Contributions**

J. L. conceived the idea of nonreciprocal field transformation and initial design, J. L. and X. W. performed numerical verification, X. W., C. C., and X. Z. established the experiment setup. X. W. made the measurements. All authors contributed to writing the manuscript. J. L. and N. P. manage the project.

**Competing Financial Interests**

Authors declare no competing interests.

**Data availability**

The data that support the plots within this paper and other findings of this study are available from the corresponding author upon reasonable request.